\documentclass[preprint2]{aastex}
\usepackage{graphicx}
\usepackage{rotating}
\usepackage[section]{placeins}

\begin{document}
\newcommand{\hr}{H$_R$\ }
\newcommand{\hv}{H$_V$\ }
\newcommand{\hjpl}{H$_V$(JPL)\ }
\newcommand{\hmpc}{H$_V$(MPC)\ }
\newcommand{\hmb}{H$_{MB}$\ }
\newcommand{\hrt}{H$_V$(ours)\ }
\newcommand{\heso}{H$_{ESO}$\ }
\newcommand{\qltforty}{q $<$ 40 AU\ }
\newcommand{\qgtforty}{q $>$ 40 AU\ }
\newcommand{\aital}{{\it a}\ }
\newcommand{\eital}{{\it e}\ }
\newcommand{\iital}{{\it i}\ }
\newcommand{\sigmac}{$\sigma_c$}
\newcommand{\sigmah}{$\sigma_h$}

\renewcommand{\arraystretch}{0.5}

\title{Colors of Inner Disk Classical Kuiper Belt Objects}

\author{W. Romanishin\altaffilmark{1}}
\affil{Physics and Astronomy Dept., University of Oklahoma, Norman, OK
73019}
\email{wromanishin@ou.edu}

\author{S. C. Tegler\altaffilmark{1}}
\affil{Physics and Astronomy Dept., Northern Arizona University,
Flagstaff, AZ 86011}
\email{Stephen.Tegler@nau.edu}

\author{G. J. Consolmagno, S.J.\altaffilmark{1}}
\affil{Vatican Observatory, Specola Vaticana, V-00120, Vatican City 
State}
\email{gjc@specola.va}

\altaffiltext{1}{Observers at the Keck 1 and Vatican
Advanced Technology telescopes}

\begin{abstract}
We present new optical broadband colors, obtained with the Keck 1
and Vatican  Advanced Technology
telescopes, for  six objects in the inner classical Kuiper 
Belt.
Objects in the inner classical Kuiper Belt 
are of interest as they may 
represent the surviving
members of the primordial Kuiper Belt that formed
interior to the current position of the 3:2 resonance with
Neptune, the current position of the plutinos,
or, alternatively, they may be objects formed at a different
heliocentric distance that were then moved to their present 
locations.
The six new colors, combined with four previously published,
show that the 10 inner belt objects with known colors form a  neutral 
clump 
and  a reddish clump
in B$-$R color.
Nonparametric statistical tests show no significant difference between
the B$-$R color distribution of the inner disk objects compared 
to the color distributions of
Centaurs, plutinos, or scattered disk objects.
However, the B$-$R color distribution of the
inner classical Kuiper belt objects {\it does} differ
significantly from the distribution of colors
in the cold (low inclination) main classical Kuiper belt.
The cold main classical objects are predominately red, while the
inner classical belt objects are a mixture of neutral
and red. 
The color difference may reveal the existence
of a gradient in the composition and /or surface processing
history in the primordial Kuiper Belt,
or indicate that the inner disk objects are not dynamically
analogous to the cold main classical belt objects.
\end{abstract}


\section{Introduction}

Ongoing discovery and orbital and physical characterization of
Kuiper belt objects (KBOs) and related outer solar system
bodies promise new insight into the formation and evolution of
our solar system.
However, the evolving dynamical picture shows that the orbits of many, 
if not most, 
objects
have been substantially changed over time.
Thus it may be difficult to study any  differences between
objects that formed at different heliocentric distances, as the
current heliocentric distances of individual objects may differ radically 
from the 
heliocentric distances at the time of their formation.

\citet{Gomes03} hypothesized a dynamical mechanism to move the
dynamically hot (high 
values of orbital
inclination (\iital) and eccentricity (\eital))
KBOs from a massive protoplanetary disk 
truncated at 
30$-$35 AU from the Sun
outward to the present Kuiper Belt region.
This mechanism would not work for the cold KBO population
(low values of orbital
inclination (\iital) and eccentricity (\eital))
\citep{SSBNMorbiAJ}.
In the Gomes model, the cold KBO population would either have formed
{\it in situ} or would have been moved by a different
mechanism.
In the context of the Nice model of the outer solar system,
a mechanism has been proposed that could move even cold KBO
objects from a disk truncated at $\sim$ 35 AU to their present
distances \citep{LevisonNice08}.
In this model, the dynamically cold KBO objects would have formed in the 
outer
regions of the massive protoplanetary disk, perhaps in the region 
around 30 AU.

The vast majority of the known 
dynamically cold classical belt objects have semimajor axes
(\aital) between 42 and 48 AU.
There are also low \eital, low \iital 
classical objects with \aital $<$ 39.4 AU.
This semimajor axis is less than 
the current position of the Neptune 3:2 resonance, 
the semimajor axis at which Pluto and
the plutino KBOs are currently found.
The population of these objects, known as inner 
classical
belt KBOs (ICKBOs), with \aital between about 36 and 39.4 AU, is much 
lower than
that of the main classical belt objects ( \aital between 42 and 48 AU).
Because of the relative dearth of known classical belt objects interior to 
the
Neptune 3:2 resonance, 
no published physical studies have specifically targeted these
objects.
This class of objects extends
the \aital range of objects  compared to the 
\aital range spanned by the main belt alone.
Thus it is 
potentially useful in the
study of any gradients of chemical or surface processing 
history with formation distance for minor bodies in the outer solar 
system.
Here we present new optical colors of a sample of inner classical
KBOs.
We compare the color distribution of these objects with
that of other classes of outer solar system objects.

\section{The Inner Classical Kuiper Belt}

A dynamical classification scheme for minor bodies in the outer solar 
system
is presented by \citet{GlaMarVan08} (hereafter GMV).
They define  inner classical belt objects as those that have 
semimajor axes (\aital)\ less than  39.4 
AU, are not Centaurs, are not scattered disk objects, and are
not in resonant orbits.
As explained in GMV, the stable inner classical belt is not disconnected 
from
the main classical belt, which contains 
objects on stable, non-resonant orbits with
\aital = 42$-$48 AU.

GMV only classify objects 
that met certain criteria for having sufficient
astrometry as of May 2006.
They list 17 as inner classical Kuiper belt objects
(hereafter ICKBO) objects and over 250  main classical Kuiper
belt objects (hereafter MCKBOs).
As more recent astrometry is available, we searched  the
continuously updated listing 
\footnote{http://www.boulder.swri.edu/$\sim$buie/kbo/kbofollowup.html}
of 
the Deep Ecliptic Survey (DES) team 
\citep{Elliot05} for possible ICKBOs in
addition to those identified in GMV.
We searched for objects 
with \aital $<$ 39.4 AU which the DES team definitively classifies as
``classical" objects based on their detailed orbital integrations.
Two additional members of the inner classical belt
(144897 and 2003 QA92) were found
in this way. 
Neither of these objects
are classified in GMV.

Of the 17 ICKBOs identified in GMV, five are classified by the
DES team as ``scattered near" objects, rather than classical
objects.
All five of these objects have orbital inclinations over 10\degr.
Thus, there is some disagreement over the classification of some
objects in the \aital $<$ 39.4 AU region.
However, all but one of the objects for which
colors are available have orbital inclinations less than 10\degr, 
and
so are probably members of the inner classical belt.
If we use the GMV classification, augmented by the 2 additional objects,
the inner classical belt has 19 known objects.
\citet{CFEPS09} estimates that the inner disk KBOs may 
have a population 10 to 20 times smaller than the main belt.

Evidence that the main classical Kuiper Belt is composed of a ``cold"
and ``hot" population has been presented by \citet{Brown01}.
There is some evidence of differences in the physical properties
between 
the
cold and hot populations (see \citet{Pei08} and references therein).
However the dividing line in inclination between the cold and hot 
populations is not sharp, and indeed a simple dividing line may not
be useful, as the two populations may overlap.
\citet{Brown01} models the two populations as separate Gaussians in
inclination, with the cold population having \sigmac \ =
2.2\degr \ and the hot  having \sigmah \ around 17\degr.
\citet{Gulbis06} use a inclination of about 5\degr \ to separate
cold (core) and hot (halo) classical objects.
However, 
\citet{Pei08} find that the colors of classical main belt objects
are uniformly red up to an inclination of 12\degr- that is, they do
not see a break in colors at 5\degr.

Our initial goal was to measure the color of ICKBOs with the lowest 
values of \iital and \eital, as these could be an extension of
the main cold classical belt towards the Sun.
If this were the case, the inner objects would extend the semimajor axis 
range of the classical belt and provide a larger range of semimajor axis
in which to look for correlations between semimajor axis and 
physical properties than the \aital range provided by the main cold 
classical belt KBOs alone.
However, new results on the structure of the Kuiper Belt obtained
from surveys analyzed with observational biases taken into account
\citep{CFEPS09}
indicate the possibility that the inner belt objects are not analogous
to the cold main belt  
but are perhaps more analogous 
to the hot classical belt objects.
Such hot objects originated at  significantly different
heliocentric distances compared to their present locations.
\citet{CFEPS09} argue, from the number of expected and observed 
inner disk objects at low \iital,
that the inner disk is most 
likely devoid
of a cold component, but this conclusion is uncertain due to
the small number of inner disk objects found so far in their
survey.

In contradiction to the \citet{CFEPS09} results, 
\citet{LykMuk07} 
specifically posit a {\it cold} inner disk.
These authors  state that cold
classical KBOs are located in the inner disk region- 37 AU $<$ a $<$ 40AU (q $<$ 37 AU)- 
as well as, of course, in the main belt region
(42 AU $<$ a $<$ 47.5 AU).

Thus, to summarize, the inner classical disk
may be 
an extension of
the cold main classical belt, or it may be a hotter component that will 
not
tell us directly about any  changes in physical properties 
as a function of formation radius in the cold population.
To the extent that optical colors can be related to formation
radii, colors of inner objects might give a hint as to the
origin of this population. 
More definite dynamical information on the origin of the inner objects 
must await
a larger sample of objects.
Also, careful modeling of the observational biases must be
taken into account \citep{CFEPS09}
to specify whether the ICKBOs are part of a dynamically hot or cold
population.

\section{Observations}

BVR imaging observations of five ICKBOs were obtained with
the Low Resolution Imaging Spectrometer (LRIS) \citep{Oke95}
on the Keck 1 telescope on Mauna Kea on 
20 September 2006.
We operated LRIS in its dual-channel imaging mode that allows
simultaneous imaging in a blue and a red channel.
We took 300 second images through a B (438 nm)
filter on the blue side, and alternated between a V (547 nm)
and an R (642 nm) filter on the red side, with a dichroic element
(D460) separating incoming light into blue and red channels.

There were some very intermittent light clouds during our
observations.
Repeatability of magnitudes in the B images indicate that the
clouds attenuated 
light by no more
than a few tenths of a magnitude at worst.
Clouds are almost true neutral absorbers in the optical
region \citep{Serkowski70, Honeycutt71}, with B$-$V and
V$-$R colors made bluer by no more than 0.01 mag for
1 mag of extinction.
Thanks to the dual-channel imaging operation mode, our colors should not 
be affected to any
significant extent by the presence of these thin clouds.
Observations of Landolt standard stars \citep{Landolt92} 
were used to derive
color transformation equations to the
Johnson- Kron- Cousins system, as in our previous
observations with this instrument 
\citep{TegRom03}.
Nine high quality standard stars from 
three Landolt fields 
(SA 110, 
PG 2213$-$006 and SA 95)
were used to derive transformation equations between
instrumental and system colors.
These transformation equations were indistinguishable,
within errors, to those shown in \citet{TegRom03}.
The solar color in this system is B$-$R = 0.99 \citep{Pei08}.

New colors of 119951 reported here were obtained
with  B (450 nm), V (550 nm), and R (650 nm)
glass filters in front of a 2048 $\times$ 2048 pixel
CCD camera at the f/9 aplanatic Gregorian focus of
the 1.8-m Vatican Advanced Technology Telescope
(VATT; the Alice P. Lennon telescope and Thomas J.
Bannan facility) on Mt. Graham, Arizona.
We binned the 15 micron pixels 2 $\times$ 2,
yielding 1024 $\times$ 1024 pixels, covering 6.4 $\times$ 6.4
arcmin of the sky at 0.375 arcsec per pixel.
Observations of 119951 were obtained on 7 Jun 2005,
during a 5 night run of photometric weather.
Multiple observations of
sixteen stars in three Landolt fields (PG 1633+099, 
PG 1323$-$086  and  SA 110) were used to derive transformation 
equations.
The scatter of the individual colors of the standard stars
with respect to the equation was typically less than
0.01 mag.

Results of the new observations are reported in Table 1.
The first column gives the name of the object, if one has been
assigned, otherwise the provisional designation is listed.
The second column gives the permanent number of the object,
if one has been assigned.
Other columns give the optical BVR colors and the error
in the color ({$\sigma$} is the dispersion in the individual color
measurements and {\it n} the number of independent measurements).
V magnitudes are quoted to only a tenth of a magnitude,
as thin clouds may have
affected these measurements.
Also, the time span over which each objects was observed was between
1 and 2 hours, not long enough to get a
proper rotational magnitude sampling for objects with substantial
lightcurve variations.
The last three columns give the semimajor axis (\aital),
eccentricity (\eital) and inclination (\iital).

We also searched for previously published colors
of ICKBOs, primarily using the MBOSS online database
\footnote{http://www.sc.eso.org/$\sim$ohainaut/MBOSS/ }
\citep{HaiDel02}.
Four additional objects were found, and for convenience their
BVR colors are listed in Table 1.

Although the sample of 10 colors seems small, it is about half
of the total known population of ICKBOs.
The remaining objects are small and mostly very faint,
and will require significant time on very large telescopes to obtain 
good colors.
A histogram of the B$-$R colors of the objects in Table 1 is shown in 
Figure 1.

\section{Colors of Inner Disk Objects}

The histogram of B$-$R colors for the ICKBOs shown in Figure 1
certainly suggest the possibility of a bimodal
color distribution.
However,
the ``dip test" \citep{Har85,Pei03} does not indicate a 
statistically 
significant bimodal signal for the present sample (~50\% rejection of
unimodal hypothesis).
The ``dip test" may, in a sense, be too restrictive 
when the question 
is whether or not 
two classes of
objects are present.
It {\it is} possible to combine two distinct classes of objects
and get a combined distribution with a {\it single} mode \citep{Zhu07}.
The definitive discussion of unimodality or bimodality of
the inner disk object colors must wait until a
larger sample of colors is available.

We use two different nonparametric or distribution-free statistical
tests to compare the B$-$R color distributions
of ICKBOs with samples of other classes of outer solar system objects.
The Wilcoxon rank-sum test,
also known as the Mann- Whitney  test,
calculates the probability that two populations 
have the same mean, so could arise from the same parent population
\citep{AldRoe77,Langley71}.
This test  compares the sum of the rankings of the 
values of the individual samples
in a pooled, ordered union of the two samples.
The test is useful for comparing samples which are not
distributed normally, as the test requires no assumption of
normal or Gaussian distributed populations.
We also use the more familiar Kolmogorov- Smirnov (or K-S) test
\citep{PressNumRec}. 
We use 
implementations of these tests found in the 
statistical package R \footnote{http://www.R-project.org}.

We first compare the B$-$R colors  of the 10 ICKBOs with B$-$R colors  
to samples of 26 Centaurs and 17 scattered disk 
object (SDOs) 
from \citet{SSBNcen}.
Next we compare the ICKBO sample to the sample
of 41 plutinos with B$-$R colors \citep{RomTeg07}.
Histograms of the B$-$R colors of the samples of
Centaurs, SDOs and plutinos used are shown in Figure 1.
The Wilcoxon test shows that the probability that the distributions 
of 10 ICKBOs and the 26 Centaurs have the same mean color is 90\%.
For the ICKBOs and the plutinos the probability is 36\%, and
for the ICKBOs and the SDOs the probability is 80\%.
Thus, we find no evidence of any statistically significant difference 
between the average color of 
ICKBO and Centaur or SDO B$-$R color distributions.
While there is a greater difference between the ICKBO
and the plutino B$-$R color distributions, it is only at the one sigma
level and so  is not statistically significant.

If the inner objects are an extension of the cold main belt population,
then any difference in color distribution could indicate a radial
gradient in physical properties.
As discussed earlier, an alternative view is that the inner disk
objects are not part of the cold classical disk.
If this is so, the
objects we have studied are just the low \iital and \eital 
members of a broader inclination distribution akin to the hot main belt
objects.
Main belt KBOs, particularly those that have low \eital and \iital,
are very predominately red \citep{TegRom00,TruBro02,Dor02},
with  B$-$R colors typically around 1.7.
From the results of \citet{Pei08},
hereafter PLJ,
we first use B$-$R colors of objects up to \iital of 12\degr \ 
as the cold main belt population distribution to compare to the
ICKBO colors.

Our first comparison is between the B$-$R colors of the 10 objects in Table 1 
and
the 46 objects from PLJ with \iital up to 12\degr.
The Wilcoxon test shows that the probability that the distributions
of B$-$R colors of 10 ICKBOs and the 46 objects from PLJ 
have the same average color  is 0.1\%.
The 
Kolmogorov-Smirnov two-sample test
applied to the same samples also 
indicates  the probability that the distributions
of B$-$R colors of 10 ICKBOs and the 46 objects from PLJ 
have the same average color is 0.1\%.

Next we restrict the samples to \iital $<$ 5\degr, as several groups
propose this as a boundary between cold and hot classical populations
\citep{Gulbis06,Noll2pop08}.
This leaves  7 objects from Table 1 and 34 objects from PLJ.
The Wilcoxon test shows that the probability that the distributions
of the 7 ICKBOs and the 34 objects from PLJ 
have the same average color is 1\%.
The
Kolmogorov-Smirnov two-sample test
applied to the same samples
indicates  probability that the distributions
of 7 ICKBOs and the 34 objects from PLJ 
have the same average color is 2\%.
The decrease in sample size of the \iital $<$ 5\degr \ samples
presumably accounts for the slight increase in probability for the
smaller samples.

\section{Discussion}

When we began this particular project,
we thought that the low \eital and \iital inner belt objects were a
natural sunward extension of the cold classical population found
between 42 $<$ a $<$ 48 AU at low \eital and \iital . Our original aim 
was simply to
measure colors for a sample of these objects to compare the colors with
the cold main belt objects. 
As discussed in Section II,
the dynamical status of the low
\eital , \iital inner disk objects is a major question. The importance 
of 
these
objects is only now being recognized; for example, the Nice model of
\citet{LevisonNice08} said nothing specifically about the inner 
classical
disk.

Much as we would like to have a definite answer to the precise
relationship between the inner-classical KBOs and the cold-classical
KBOs, neither our color data nor the dynamical models can resolve this
issue at present. One of the main issues is the dynamical origin of the
inner classical objects, as discussed in section II above. So long as
this issue remains unsettled, we are left with admittedly frustrating
result of two different possible ``conclusions" for how our colors may
relate these objects to the other populations. Though of course more
color data are always welcome, and could perhaps provide a stronger
statement about the connection, or lack of connection, with the plutino
population, that will have to wait for further discoveries of brighter
inner classical objects amenable to measuring colors. In the meanwhile,
we await further developments from the dynamicists to propose a possible
resolution to this dilemma.

\section{Conclusions}

The B$-$R color distribution of a sample of 10 inner disk Kuiper Belt 
objects is shown to be not inconsistent with the color distribution
of samples
of plutinos, Centaurs and scattered disk objects.
%The current inner disk sample does not show statistically significant 
%evidence for having a bimodal color distribution, as do the Centaurs.
The current inner disk sample has both red and neutral colored objects,
reminiscent of the Centaurs, 
which have a bimodal B$-$R color distribution.
However, 
we cannot claim that the ICKBOs have a statistically significant bimodal
signal with the present data.

The average B$-$R color of the ICKBOs are inconsistent, at the 98\% level
or higher, depending on specific sample and statistical test used, 
with the colors of a sample of cold (low inclination) classical
KBOs with semimajor axes between 42 and 48 AU.
Possible conclusions of this are:

1) The inner disk objects we observed, even though they are of
low \eital and \iital, are members of a population analogous to the 
{\it hot} 
classical KBOs.
The hot classical KBOs do not show a predominately red optical
color distribution, as do the cold classical objects.

2) If the inner disk objects are in fact members of a cold inner disk 
population which is a sunward continuation of the cold classical
objects between 42 and 48 AU, then there is a radial color
gradient in the colors of this cold disk population of KBOs.

\acknowledgements 

We thank the NASA Planetary Astronomy program for financial support of
this research and the NASA Keck and Vatican
Observatory telescope allocation committees for consistent allocation
of telescope time.
We thank Brett Gladman and J.J. Kavelaars for discussions of the inner 
classical 
belt which helped motivate this work.

\vfil\eject

\vfil\eject

\begin{deluxetable}{lllccccccccc}
\tablewidth{0pt}
\tablecaption{New and Existing Colors for Inner Disk Objects}
\tablehead{
\colhead{Object} & 
\colhead{Number} & 
\colhead{B-R}  & \colhead{${\sigma \over \sqrt{n}}$} &
\colhead{B-V}  & \colhead{${\sigma \over \sqrt{n}}$} &
\colhead{V-R}  & \colhead{${\sigma \over \sqrt{n}}$} & 
\colhead{V} &
\colhead{a} &
\colhead{e} &
\colhead{i}
}

\startdata
\sidehead{Previously unpublished colors:}
2001 QT322\tablenotemark{a} & 135182 & 1.24 & 0.06 & 0.71 & 0.06 & 0.53 & 
0.12 & 23.9  & 37.1 & 0.017 & 1.8  \\
2002 KX14\tablenotemark{b} & 119951 & 1.66 & 0.04 & 1.05 & 0.03 & 0.61 & 
0.02 & 20.88 & 39.0 & 0.042 & 0.4 \\
2003 QA92\tablenotemark{a} & & 1.67 & 0.02 & 1.04 & 0.03 & 0.63 & 0.04 & 
22.7 & 38.0 & 0.061 & 3.4 \\
2003 QQ91\tablenotemark{a} & & 1.18 & 0.05 & 0.67 & 0.06 & 0.51 & 0.08 & 
24.0 & 38.7 & 0.074 & 5.4 \\
2003 YL179\tablenotemark{a} & & 1.18 & 0.05 & 0.67 & 0.06 & 0.51 & 0.08 & 
24.0 & 38.9 & 0.005 & 2.5 \\
2004 UX10\tablenotemark{a} & 144897 & 1.53 & 0.02 & 0.95 & 0.02 & 0.58 & 
0.03 & 20.9 & 38.9 & 0.039 & 9.5 \\
\sidehead{Previously published colors:}
1998 SN165\tablenotemark{d} & 35671 & 1.16 & 0.12 & 0.71 & 0.10 & 0.44 & 
0.08 &  & 38.2 & 0.042 & 4.6 \\
1998 WV24\tablenotemark{c} & & 1.27 & 0.03 & 0.77 & 0.01 & 0.50 & 0.03 & & 39.2 & 0.044 & 1.5 \\
Rhadamanthus\tablenotemark{d} & 38083 & 1.18 & 0.11 & 0.65 & 0.09 & 0.53 & 
0.07 & & 38.8 & 0.155 & 12.8 \\
1999 OJ4\tablenotemark{d} & & 1.77 & 0.17 & 1.10 & 0.16 & 0.67 & 0.07 & & 38.2 & 0.028 & 4.0 \\ 
\enddata
\tablenotetext{a}{Observed with Keck 1 telescope 20 Sep 2006}
\tablenotetext{b}{Observed with Vatican Advanced Technology Telescope 7 
Jun 2005}
\tablenotetext{c}{Tegler and Romanishin 2000}
\tablenotetext{d}{Hainaut 2007}
\end{deluxetable}

\clearpage

\begin{figure}
       \includegraphics{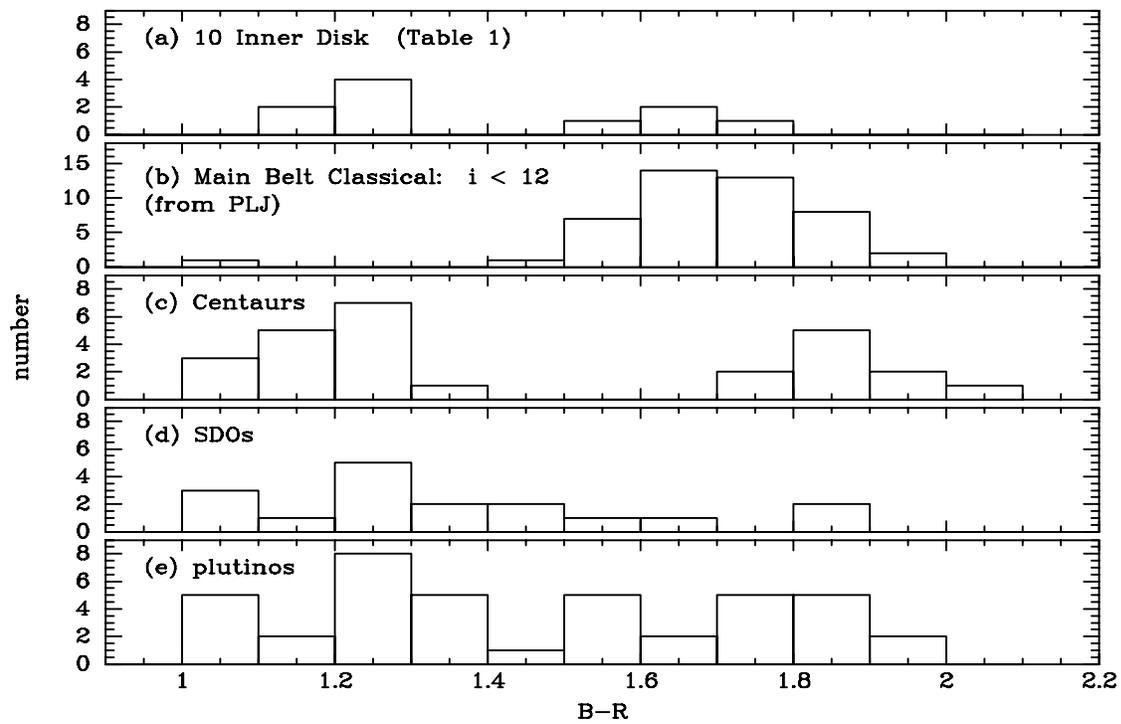}
       \caption{Histograms of B$-$R colors of 10 inner Kuiper
Belt objects and samples of other classes of objects as discussed in 
text.}
\end{figure}

\end{document}